\documentclass[%
 aip,
 amsmath,amssymb,
 reprint,%
]{revtex4-1}

\usepackage{graphicx}
\usepackage{dcolumn}
\usepackage{bm}

\usepackage[utf8]{inputenc}
\usepackage[T1]{fontenc}
\usepackage{mathptmx}
\DeclareSymbolFont{epsilon}{OML}{ntxmi}{m}{it}
\DeclareMathSymbol{\epsilon}{\mathord}{epsilon}{"0F}
\usepackage{etoolbox}
\usepackage{xcolor}
\usepackage{physics}
\usepackage{mathtools}
\usepackage[normalem]{ulem}
\newcommand{\reals}{\mathbb{R}}

\newcommand{\transpose}[1]{{#1}^{\top}}
\newcommand{\blf}[2]{\left\langle #1, #2\right\rangle}

\newcommand{\figref}[1]{Fig.~\ref{#1}}
\newcommand{\Ga}{\Gamma_{\mathrm{a}}}
\newcommand{\sip}[2]{\left[#1,#2\right]}

\makeatletter
\def\@email#1#2{%
 \endgroup
 \patchcmd{\titleblock@produce}
  {\frontmatter@RRAPformat}
  {\frontmatter@RRAPformat{\produce@RRAP{*#1\href{mailto:#2}{#2}}}\frontmatter@RRAPformat}
  {}{}
}%
\makeatother
\begin{document}



\title{Phase reduction of reaction-diffusion systems with delay} 

\author{Ayumi Ozawa}
\email{ozawaa@jamstec.go.jp}
\homepage{https://researchmap.jp/ayumi_ozawa}
\affiliation{
Center for Mathematical Science and Advanced Technology, Japan Agency for Marine-Earth Science and Technology, Yokohama 236-0001, Japan
}

\author{Yoji Kawamura}

\affiliation{
Center for Mathematical Science and Advanced Technology, Japan Agency for Marine-Earth Science and Technology, Yokohama 236-0001, Japan
}

\date{\today}

\begin{abstract}
    \footnotesize{The following article has been submitted to Chaos: An Interdisciplinary Journal of Nonlinear Science. After it is published, it will be found at \url{https://publishing.aip.org/resources/librarians/products/journals/}.}\vspace{1em}\\
    \normalsize
    We develop a phase reduction method for reaction-diffusion systems with a discrete delay.
    On the basis of the recent developments in the phase reduction theory for infinite-dimensional systems, we introduce a bilinear form tailored to spatially extended systems involving a discrete delay. 
    By solving the adjoint equation associated with the bilinear form, we obtain the phase sensitivity function, which quantifies the shift of the phase in response to a given perturbation.
    The theory is verified numerically with the use of the Schnakenberg system with a discrete delay in one spatial dimension.
    We further demonstrate the utility of the theory by optimizing the interaction between a pair of the Schnakenberg systems, with the use of the phase equation, for maximizing the stability of in-phase synchronization. 
    This study serves as a step towards establishing a theory for analyzing oscillatory systems that involve both spatial degrees of freedom and delay.
\end{abstract}

\maketitle

\begin{quotation}
     Oscillatory patterns arise widely in natural and engineered systems 
    and are often modeled with the use of partial differential equations involving delays~\cite{Wu1996, SeirinLee2010,Gaffney2006,Cessi2000, Gallego2001}.
    Theoretical tools for analyzing such equations are essential for understanding, controlling, and optimizing spatiotemporal oscillations, but they remain limited, particularly for systems far from bifurcation points. 
    Here, we develop an adjoint-based phase reduction method~\cite{Kuramoto1984, Brown2004, Ermentrout1991a, izhikevich07, Nakao2014, Kawamura2008,Kawamura2013,Godavarthi2023, Kotani2012, Novicenko2012} for a class of delay partial differential equations, namely reaction–diffusion systems with a discrete delay.
    The proposed method yields a phase equation that describes the response of a limit-cycle oscillation to weak perturbations. 
    Thus, this study facilitates systematic analysis of rhythmic phenomena, such as synchronization, in spatially extended systems with delay,
     as demonstrated for the one-dimensional Schnakenberg system with a discrete delay.
\end{quotation}

\section{Introduction}
\label{sec:introudction}
Dynamical systems that involve delay are employed in a wide range of disciplines, including biology~\cite{Mackey1977,Novak2008,Glass1988,Yoshioka-kobayashi2020}, ecology~\cite{Hutchinson1948,May1974,Yang2023}, chemistry~\cite{Epstein1990,Murakami2025}, climatology~\cite{Minobe2004,Cessi2000}, and engineering~\cite{Minorsky.1922, Brackstone1999,Negash2023}, and they often exhibit oscillation~\cite{Hale1993, Smith2011, Erneux2009,Atay2010,Wu1996}.
For example, cell-cell communication through signaling molecules involves numerous biochemical processes and may be effectively represented using delayed coupling~\cite{SeirinLee2010,Gaffney2006,Yoshioka-kobayashi2020}. 
Mathematical analyses indicate that, for a range of delay lengths, desirable or undesirable collective oscillations may arise among the cell population~\cite{SeirinLee2010,Gaffney2006,Yoshioka-kobayashi2020}. 
Also, the time required for a wave to propagate across the ocean serves as a source of memory in the climate system, and this memory effect may be mathematically described as distributed delays~\cite{Smith2011} in the governing equation. 
Such delays are proposed to be responsible for oscillatory dynamics on the decadal time scale in the climate system~\cite{Cessi2000}.

Among various delayed dynamical systems, partial differential equations (PDEs) involving delay are used when the spatial degrees of freedom should be taken into account~\cite{SeirinLee2010,Gaffney2006,Cessi2000, Wu1996,Gallego2001}.
While both spatial degrees of freedom and delay make such a system infinite dimensional, the onset of the oscillation can be analyzed by using bifurcation theory~\cite{Wu1996,Faria1995,Jiang2019}.
Far from bifurcation points, however, it remains difficult to analyze how such oscillatory patterns respond to perturbations such as external forcing, noise, and interactions with other oscillatory systems. 
This limitation restricts our understanding of the dynamical behavior of interacting systems with both delays and spatial degrees of freedom, such as synchronous variations of sea-surface temperatures across distant currents~\cite{Kohyama2021}.

Phase reduction theory has been developed for ordinary differential equations~\cite{Kuramoto1984,Brown2004,Ermentrout1991a,izhikevich07}, some classes of partial differential equations~\cite{Nakao2014,Kawamura2008,Kawamura2013,Godavarthi2023}, and delay differential equations~\cite{Kotani2012,Novicenko2012}, and it has been utilized to understand the response of oscillatory systems to a perturbation.
The theory reduces a given system to an ordinary differential equation for a scalar variable, called the phase.
In the resulting phase equation, the effect of the perturbation is quantified with the use of a function called the phase sensitivity function; the shift of the instantaneous frequency due to the perturbation is given by the inner product between the phase sensitivity and the perturbation.
Thus, the phase reduction method fosters theoretical investigation of oscillatory systems exposed to perturbation.
However, phase reduction theory for delay partial differential equations has not yet been developed.

In this paper, 
we take a step toward developing the phase reduction theory for partial differential equations with delay by formulating a phase reduction method for reaction-diffusion systems with a discrete delay in the reaction part. 
Methodologically, our approach can be regarded as an extension of the adjoint method,
which has proven useful for the reduction of infinite-dimensional systems~\cite{Nakao2014,Novicenko2012,Kotani2012,Kawamura2011}. 
We extend the method by introducing an appropriate bilinear form and deriving an adjoint equation.
The proposed theory is numerically validated by using the Schnakenberg system with delay as an example.
Furthermore, we demonstrate the usefulness of the theory by optimizing the interaction between a pair of the Schnakenberg systems for mutual synchronization.

This paper is organized as follows. In Sec.~\ref{sec:phase_reduction_theory}, we describe the class of delayed reaction-diffusion systems under consideration and formulate a phase reduction method for this class of systems. 
In Sec.~\ref{sec:numerical_validation}, we validate the theory through numerical simulations, using the Schnakenberg system with a delay in one spatial dimension as an example. 
In Sec.~\ref{sec:optimal_filter}, we demonstrate the utility of the proposed phase reduction method by maximizing the stability of the in-phase state in a pair of Schnakenberg systems with delays through analysis of the corresponding phase equations.
Finally, in Sec.~\ref{sec:concluding_remarks}, we provide a summary and discussion on future directions.

\section{Theory}
\label{sec:phase_reduction_theory}
In this section, we introduce a class of reaction-diffusion equations with a discrete delay and obtain a phase equation. 
The phase sensitivity function, which plays a key role in the phase equation, is obtained by solving an adjoint equation of the linearization of the system around an oscillatory solution.
\subsection{Reaction-diffusion equation with delay}
We consider the following reaction-diffusion system with a discrete delay and perturbation defined on an open region in the $M$-dimensional Euclidean space, $\Omega \subset \reals^M$, with smooth boundary $\partial \Omega$:
\begin{align}
    \pdv{\bm{u}(\bm{r}, t)}{t} 
    &= \bm{R}\left(\bm{u}(\bm{r}, t), \bm{u}( \bm{r}, t-\tau), \bm{r}\right) \notag \\
    &~~~~+ \hat{D} \nabla^2 \bm{u}(\bm{r}, t) + \epsilon \bm{p}(\bm{r}, t), \label{eq:RD}
\end{align}
where $\bm{u}(\bm{r}, t) \in \reals^{N}$ represents the density of components at time $t$ and position $\bm{r} \in \Omega$, $N$ is the number of components, $\bm{R}:\reals^{N}\times \reals^{N} \times \Omega \to \reals^{N}$ describes local reactions at $\bm{r}$, $\tau > 0$ is a constant representing a delay, and $\hat{D}$ is the $N \times N$ diffusion matrix. 
The reaction $\bm{R}$ may in general be heterogeneous over $\Omega$, although we choose a homogeneous case when illustrating examples in Secs.~\ref{sec:numerical_validation} and \ref{sec:optimal_filter}.
The term $\epsilon \bm{p}(\bm{r}, t)$ describes perturbation, where $\epsilon$ is a sufficiently small constant. 
For simplicity, we impose the zero Neumann boundary condition:
\begin{align}
    \mathbf{n}(\bm{r})\cdot \nabla u_n(\bm{r}, t) = 0 \text{ on $\partial \Omega$},
    \label{eq:zero_neumann}
\end{align}
where $\mathbf{n}(\bm{r})$ is the outward pointing normal to $\partial \Omega$ at $\bm{r}$ and $u_n(\bm{r}, t)$ $(n=1,2,\dots,N)$ is the $n$th component of $\bm{u}(\bm{r}, t)$.

We assume that, in the absence of the perturbation $\epsilon \bm{p}(\bm{r}, t)$, an oscillatory pattern arises. Namely, for $\epsilon = 0$, Eq.~\eqref{eq:RD} has a stable limit cycle solution $\bm{\chi}(\bm{r}, t)$ with period $T$: $\bm{\chi}(\bm{r}, t+T) = \bm{\chi}(\bm{r}, t)$ for any $\bm{r} \in \Omega$ and $t \in \reals$. 
The solution orbit $\gamma \coloneqq \left\{ \left.\bm{\chi}(\bm{r}, t)\right| \bm{r} \in \Omega, t \in \reals \right\}$ can be parameterized in terms of a variable that evolves at the constant frequency
\begin{align}
 \omega \coloneqq 2\pi/T.   \label{eq:nat_freq} 
\end{align}
Namely, if we define $\bm{\chi}_{\phi}(\bm{r}) \coloneqq \bm{\chi}(\bm{r}, \phi/\omega)$, where the phase variable $\phi$ satisfies
\begin{align}
    \dv{\phi(t)}{t} = \omega \label{eq:phase_eq_noperturb}
\end{align}
and $\phi(0)=0$, we have $\bm{\chi}_{\phi(t)}(\bm{r}) = \bm{\chi}(\bm{r}, t)$.
Our goal in the following subsection is to extend Eq.~\eqref{eq:phase_eq_noperturb} to the case where the solution deviates from $\gamma$ due to the perturbation $\epsilon \bm{p}(\bm{r}, t)$.

\subsection{An adjoint method}
\label{subsec:adjoint}
In the absence of delay or spatial degrees of freedom, 
the deviation of the instantaneous frequency from the natural frequency $\omega$ due to a perturbation is given by the inner product of the phase sensitivity and the perturbation~\cite{Brown2004, Nakao2014, Novicenko2012, Kotani2012, Ermentrout1991a}. 
Among several approaches for obtaining the phase sensitivity function~\cite{Brown2004}, the adjoint method involves finding a periodic solution of the adjoint equation associated with the linearized equation of the original system around the limit cycle~\cite{Brown2004, Nakao2014, Novicenko2012, Kotani2012, Ermentrout1991a}.
Here, we aim to formulate an adjoint method applicable to Eq.~\eqref{eq:RD}.

To this end, we first linearize Eq.~\eqref{eq:RD} around the limit-cycle solution $\bm{\chi}$ for $\epsilon = 0$:
\begin{align}
    \pdv{\bm{v}(\bm{r}, t)}{t} &= \hat{L}(\bm{r}, t)\bm{v}(\bm{r}, t),
    \label{eq:linearized}
\end{align}
where 
\begin{align}
    \bm{v}(\bm{r}, t) &\coloneqq \bm{u}(\bm{r}, t) - \bm{\chi}(\bm{r}, t), \notag \\
    \hat{L}(\bm{r}, t)\bm{v}(\bm{r}, t)
    &= \hat{R}_1(\bm{r}, t)\bm{v}(\bm{r}, t) +\hat{R}_2(\bm{r}, t)\bm{v}(\bm{r},t - \tau) \notag \\
    &~~~~+\hat{D} \nabla^2 \bm{v}(\bm{r}, t),
    \label{eq:define_L}
\end{align}
and the matrix-valued function $\hat{R}_i~(i=1,2)$ is defined to be the derivative of $\bm{R}$ with respect to the $i$th argument evaluated at the limit cycle. Namely, 
\begin{subequations}
\begin{align}
    \hat{R}_1(\bm{r}, t) = 
     \left. \pdv{\bm{R}(\bm{u}, \bm{u}_\tau, \bm{r})}{\bm{u}}
      \right|_{\substack{\bm{u} = \bm{\chi}(\bm{r},~t) \\ 
      \bm{u}_\tau = \bm{\chi}(\bm{r},~t-\tau)}},\\ 
    \hat{R}_2(\bm{r}, t) = 
     \left. \pdv{\bm{R}(\bm{u}, \bm{u}_\tau, \bm{r})}{\bm{u}_\tau} 
      \right|_{\substack{\bm{u} = \bm{\chi}(\bm{r},~t) \\ \bm{u}_\tau = \bm{\chi}(\bm{r},~t-\tau)}}.
\end{align}    
\end{subequations}
We again impose the zero Neumann boundary condition for the consistency with the original system:
\begin{align}
     \mathbf{n}(\bm{r})\cdot \nabla v_n(\bm{r}, t) = 0 \text{ on $\partial \Omega$},
     \label{eq:zero_neumann_linarized}
\end{align}
where $v_n(\bm{r}, t)$ $(n=1,2,\dots,N)$ is the $n$th component of $\bm{v}(\bm{r}, t)$.
Note that the operator $\hat{L}$ is $T$-periodic and $\hat{L}(\bm{r}, t+ T)\bm{v}(\bm{r}, t+T) = \hat{L}(\bm{r}, t)\bm{v}(\bm{r}, t) $ for a periodic solution $\bm{v}(\bm{r}, t)$.

Equation \eqref{eq:linearized} can be formulated as a functional differential equation in a certain infinite-dimensional Banach space as below~\cite{Wu1996,Schumacher1985}. 
Let $\bm{v}^{(t)}$ denote the element of $C([-\tau, 0]; C(\Omega;\reals^{N}))\eqqcolon \mathcal{C}$ given by $\bm{v}^{(t)}(\sigma)(\bm{r}) = \bm{v}(\bm{r}, t + \sigma)$,
where $C(X;Y)$ denotes, given Banach spaces $X$ and $Y$, the Banach space of continuous $Y$-valued functions on $X$ with supremum norm. Then Eq.~\eqref{eq:linearized} determines the evolution of $\bm{v}^{(t)}$ on $\mathcal{C}$. 

As an analogy of the case of delay differential equations~\cite{Simmendinger1999}, we introduce the dual space $\mathcal{C}^*$ of $\mathcal{C}$ as $ \mathcal{C}^* = C([0,\tau]; C(\Omega; {\reals^{N}}^*))$, where ${\reals^{N}}^{*}$ is the space of row vectors of $N$ real components.
We also define a time-dependent bilinear form for the elements of $\mathcal{C}^*$ and $\mathcal{C}$ as 
\begin{align}
    &\blf{\bm{w}(\sigma)(\bm{r})}{\bm{s}(\sigma)(\bm{r});t} \notag\\
    &\coloneqq 
    \sip{\bm{w}(0)(\bm{r})}{\bm{s}(0)(\bm{r})} \notag \\
    &~~~~+ \int_{-\tau}^{0} 
            \sip{\bm{w}(\tau+\sigma)(\bm{r})}
                {\hat{R}_2(\bm{r}, t+\tau+\sigma)\bm{s}(\sigma)(\bm{r})}
        \dd \sigma,\label{eq:blf} \\
    &\bm{w} \in \mathcal{C}^*, ~\bm{s} \in \mathcal{C}, \notag
\end{align}
where $\sip{\cdots}{\cdots}$ is the spatial integral of the product of the two vector-valued functions over $\Omega$:
\begin{align}
    &\sip{\bm{w}(\sigma)(\bm{r})}{\bm{s}(\sigma)(\bm{r})} 
    \coloneqq \int_\Omega \bm{w}(\sigma)(\bm{r})\bm{s}(\sigma)(\bm{r}) \dd \bm{r}.
    \label{eq:sip}
\end{align}
The bilinear form \eqref{eq:blf} is constructed by incorporating the spatial degrees of freedom into a bilinear form that is used in the development of the Floquet theory for delay differential equations~\cite{Simmendinger1999}.
We then require the dual $\bm{q}^{(t)}(\sigma)(\bm{r}) \in \mathcal{C}^*$ of $\bm{v}^{(t)}(\sigma)(\bm{r})$ to satisfy 
\begin{align}
    \dv{}{t}\blf{\bm{q}^{(t)}(\sigma)(\bm{r})}{\bm{v}^{(t)}(\sigma)(\bm{r});t} = 0,
    \label{eq:constant_blf}
\end{align}
and seek for a function $\bm{q}(\bm{r},t)$ such that $\bm{q}^{(t)}(\sigma)(\bm{r}) = \bm{q}(\bm{r},t+\sigma)$. As detailed in Appendix \ref{sec:adjoint}, Eq.~\eqref{eq:linearized} and Eq.~\eqref{eq:constant_blf} yield the following adjoint equation:
\begin{align}
    \pdv{\bm{q}(\bm{r}, t)}{t} &= 
        - \bm{q}(\bm{r}, t) \hat{R}_1 (\bm{r}, t)
        - \bm{q}(\bm{r}, t+\tau)\hat{R}_2 (\bm{r}, t+\tau)  \notag \\
        &~~~~~- \nabla^2 \bm{q}(\bm{r}, t) \hat{D}, ~~~~\bm{r} \in \Omega,\label{eq:adjoint} \\
        \mathbf{n}(\bm{r})&\cdot \nabla q_n(\bm{r}, t) = 0, ~~~~~~\bm{r} \in \partial \Omega,
        \label{eq:zero_neumann_adjoint}
\end{align}
where $q_n(\bm{r}, t)$ $(n=1,2,\dots,N)$ is the $n$th component of $\bm{q}(\bm{r}, t)$.
We assume that the adjoint equation \eqref{eq:adjoint} has a periodic solution and let $\bm{Q}(\bm{r}, t)$ denote the solution.

Note that the bilinear form of 
\begin{align}
    \bm{Q}^{(t)}(\sigma)(\bm{r}) \coloneqq \bm{Q}(\bm{r}, t+\sigma) \label{eq:qt_history}
\end{align}
and any nonzero Floquet mode of Eq.~\eqref{eq:linearized} vanishes.
Namely, if we assume that Eq.~\eqref{eq:linearized} has a solution of the form 
\begin{align}
    \bm{v}_\lambda(\bm{r}, t) = e^{\lambda t} \bm{p}_{\lambda}(\bm{r}, t),
    \label{eq:v_lambda}
\end{align}
 where $\lambda \neq 0$ is a constant and $\bm{p}_{\lambda}(\bm{r}, t)$ is $T$-periodic with respect to time $t$, then
\begin{align}
    \blf{\bm{Q}^{(t)}(\sigma)(\bm{r})}{\bm{v}^{(t)}_\lambda(\sigma)(\bm{r}); t} = 0, \label{eq:orthogonal}
\end{align}
where $\bm{v}^{(t)}_\lambda(\sigma)(\bm{r}) \coloneqq \bm{v}_{\lambda}(\bm{r},t+\sigma)$.

Since the zeroth Floquet mode, or the periodic solution, of Eq.~\eqref{eq:linearized} does not decay with time, it can be interpreted as the shift of the phase from that of $\bm{\chi}(\bm{r},t)$. Hereafter we call the zeroth Floquet mode phase mode.
In contrast, other Floquet modes can be interpreted as the deviation from the limit-cycle because it decays with time. 
Thus, $\bm{Q}^{(t)}(\sigma)(\bm{r})$ extracts only the phase mode in the sense that Eq.~\eqref{eq:orthogonal} holds.
See Appendix \ref{sec:zero_eigenfunction} for the derivation of Eq.~\eqref{eq:orthogonal}.

One can show that the phase mode can be chosen to be $\bm{U}(\bm{r}, t) \coloneqq \partial \bm{\chi}(\bm{r}, t)/\partial t$ by inserting $\bm{u}(\bm{r},t)=\bm{\chi}(\bm{r}, t)$ and $\epsilon=0$ into Eq.~\eqref{eq:RD} and differentiate it with respect to $t$. As in the conventional adjoint methods~\cite{izhikevich07}, we use this solution to normalize $\bm{Q}(\bm{r}, t)$ as follows:
\begin{align}
    \blf{\bm{Q}^{(t)}(\sigma)(\bm{r})}{\bm{U}^{(t)}(\sigma)(\bm{r}); t} = \omega,
    \label{eq:normalization}
\end{align}
where $\bm{U}^{(t)}(\sigma)(\bm{r}) \coloneqq \bm{U}(\bm{r}, t + \sigma)$ and $\omega$ is defined by Eq.~\eqref{eq:nat_freq}.  As with $\bm{\chi}$, we can write $\bm{Q}(\bm{r}, t)$ and $\bm{U}(\bm{r}, t)$ in terms of the phase $\phi$.
We define $\bm{Q}_\phi$ and $\bm{U}_\phi$ by setting 
\begin{align}
    \bm{Q}_\phi(\bm{r}) \coloneqq \bm{Q}(\bm{r}, \phi/\omega) \label{eq:Q_phi}
 \end{align}
and
 \begin{align}
    \bm{U}_\phi(\bm{r}) \coloneqq \bm{U}(\bm{r}, \phi/\omega) \label{eq:U_phi}, 
\end{align}
respectively. 

Finally, we extract the phase mode from the perturbation $\epsilon \bm{p}(\bm{r}, t)$ using $\bm{Q}^{(\phi/\omega)}(\sigma)(\bm{r})$. 
Note that a perturbation given at time $t$ does not affect the value of $\bm{u}$ before time $t$. Hence, as a function from $[-\tau, 0]$ to $C(\Omega;\reals^{N})$, the perturbation is written as 
\begin{align}
    \bm{p}^{(t)}(\sigma)&(\bm{r}) =
    \begin{cases}
        \bm{p}(\bm{r}, t) & (\sigma = 0) \\
        0 & (-\tau \leq \sigma < 0).
    \end{cases}
\end{align}
By calculating the bilinear form between $\epsilon \bm{p}^{(t)}(\sigma)(\bm{r})$ and $\bm{Q}^{(\phi/\omega)}$, we obtain\footnote{Although $\bm{p}^{(t)}(\sigma)(\bm{r})$ is not a smooth function in terms of $\sigma$ and thus not an element of $\mathcal{C}$, we can still calculate the right hand side of Eq.~\eqref{eq:blf} for $\bm{w}=\bm{Q}^{(\phi/\omega)}$ and $\bm{s}=\bm{p}$. 
Hence, the definition of the bilinear form Eq.~\eqref{eq:blf} is sufficient for the practical purpose, while the development of a theory that deals with a larger space than $\mathcal{C}$ remains a future issue.
}
\begin{align}
    \dv{\phi(t)}{t} 
        &= \omega + \epsilon\blf{\bm{Q}^{(\phi/\omega)}(\sigma)(\bm{r})}{\bm{p}^{(t)}(\sigma)(\bm{r}); t} \notag \\
        &= \omega + \epsilon\sip{\bm{Q}_\phi(\bm{r})}{\bm{p}(\bm{r}, t)}.
        \label{eq:phase_equation}
\end{align}
Equation \eqref{eq:phase_equation} implies that $\bm{Q}_{\phi}(\bm{r})$ quantifies the response of the phase $\phi$ to the perturbation $\epsilon\bm{p}(\bm{r}, t)$, and therefore, it serves as the phase sensitivity function~\cite{Nakao2016,Winfree1967}.

\begin{figure}[t]
    \centering
     \includegraphics[width=\columnwidth]{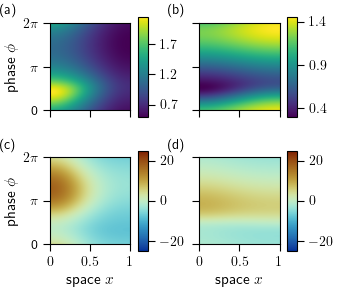}
    \caption{Limit cycle and phase sensitivity function. The $u$ and $v$ components of $\bm{\chi}_{\phi}(\bm{r})$ are plotted in (a) and (b). Also, the $u$ and $v$ components of $\bm{Q}_{\phi}(\bm{r})$ are plotted in (c) and (d).
    }
    \label{fig:lc}
\end{figure}

\section{Numerical validation}
\label{sec:numerical_validation}
In this section, we use the delayed Schnakenberg system in one spatial dimension proposed in Ref.~\cite{SeirinLee2011} as a simple example of reaction-diffusion systems with a discrete delay and numerically show that the phase equation \eqref{eq:phase_equation} captures the modulation of the rhythm in response to perturbation.
\subsection{The Schnakenberg system on a one-dimensional spatial domain}
We consider the following one-dimensional Schnakenberg system with a discrete delay defined on $\Omega = (0,1)$ with zero Neumann boundary condition~\cite{SeirinLee2011}:
\begin{subequations}
\begin{align}
        &\bm{u}(x,t) = 
        \begin{bmatrix}
            u(x,t) \\ v(x, t)
        \end{bmatrix}, \notag \\
        &\bm{R}\left(\bm{u}(x, t), \bm{u}(x, t-\tau)\right)\notag \\
         &=
         \begin{bmatrix}
             a - u(x,t) + u(x,t-\tau)^2 v(x,t-\tau)\\
             b - u(x,t-\tau)^2 v(x,t-\tau) 
         \end{bmatrix} \label{eq:schnakenberg_reaction}, \\
         \hat{D} &= 
         \begin{bmatrix}
             \varepsilon D_0 & 0 \\
            0 &D_0
         \end{bmatrix}, \label{eq:Diffusion_matrix}
\end{align}
\label{eq:schnakenberg}
\end{subequations}
where $a, b, D_0, \varepsilon$, and $\tau$ are real constants. 

The bifurcation analyses of this system imply that a spatially nonuniform stable limit-cycle solution exists for a certain range of parameters ~\cite{Jiang2019}.
We set $a = 0.1$, $b=0.9$, $D_0=0.5$, $\varepsilon=0.817$ and the delay $\tau=0.4071$, with which $u(x,t)$ and $v(x,t)$ exhibit oscillation with period $T \simeq 7.06$ as shown in Figs.~\ref{fig:lc} (a) and (b), respectively.

Using this periodic solution, we numerically integrated Eq.~\eqref{eq:adjoint} to obtain the phase sensitivity function $\bm{Q}_\phi(x)$. Its $u$ and $v$ components are shown in Figs.~\ref{fig:lc} (c) and (d), respectively. Note that, as in the conventional adjoint methods~\cite{Nakao2014}, we integrate Eq.~\eqref{eq:adjoint} backward in time for the following reason. 
The stability of the limit cycle implies that any solution of Eq.~\eqref{eq:linearized} converges to a periodic solution or zero. 
The former corresponds to shift of the phase from that of $\bm{\chi}(t)$, while the latter corresponds to the deviation from the limit cycle.
The dual of a decaying mode should diverge for the condition \eqref{eq:constant_blf} to be satisfied. 
Hence, in general, a solution of the adjoint equation \eqref{eq:adjoint} includes diverging modes. 
By integrating Eq.~\eqref{eq:adjoint} backward in time, we can expect that these modes vanish after a transient and a periodic solution is obtained.

\subsection{Phase sensitivity function}
\label{subsec:phase_sensitivity}
We now numerically show that the response of the phase to a pulse-like perturbation can be captured by using the phase sensitivity function $\bm{Q}_{\phi}(x)$.
Let us consider $\bm{p}(x, t)$ of the following form
\begin{align}
    \bm{p}(x, t) =
    \begin{cases}
        \bm{p}(x)/\Delta t & (t_0 \leq t \leq t_0 + \Delta t) \\
        0 & \text{(otherwise)}
    \end{cases},
    \label{eq:perturbation}
\end{align}
where $t_0$ is the time at which the perturbation is applied and $\Delta t$ is a sufficiently small constant.
We let $\mathrm{PRC}(\phi; \epsilon \bm{p}(x))$ denote the phase response curve. Namely, given the phase $\phi$ at time $t_0$, the value of $\mathrm{PRC}(\phi; \epsilon \bm{p}(x))$ represents the shift of the phase due to the perturbation $\epsilon\bm{p}(x, t)$.

On the one hand, $\mathrm{PRC}(\phi; \epsilon \bm{p}(x))$ can be obtained by performing numerical simulation of Eq.~\eqref{eq:RD} with Eq.~\eqref{eq:schnakenberg} and comparing the phase of the perturbed system at $t \gg t_0$ with that of the unperturbed system with the same initial condition. On the other hand, the phase equation \eqref{eq:phase_equation} implies 
\begin{align}
    \mathrm{PRC}(\phi; \epsilon \bm{p}(x))/\epsilon \simeq \sip{\bm{Q}_{\phi}(x)}{\bm{p}(x)} \label{eq:phase_response}
\end{align}
for sufficiently small $\epsilon$ and $\Delta t$.

\begin{figure}[]
    \centering
    \includegraphics{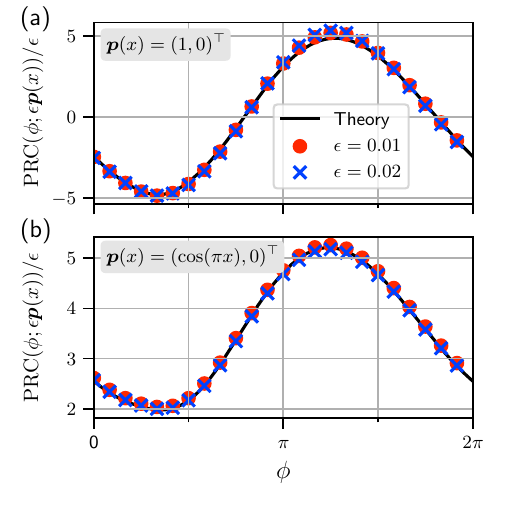}
    \caption{Phase response curve of the delayed Schnakenberg system to the perturbation (a) $\bm{p}(x)=\transpose{(1,0)}$ and (b) $\bm{p}(x)=\transpose{(\cos \pi x, 0)}$. The orange circles and blue crosses are the results of the direct numerical simulation with the perturbation strengths $\epsilon = 0.01$ and $\epsilon = 0.02$, respectively. The black solid curves are obtained from Eq.~\eqref{eq:phase_response}}
    \label{fig:PRC}
\end{figure}

As shown in \figref{fig:PRC}, the values of $\mathrm{PRC}(\phi; \epsilon \bm{p}(x))/\epsilon$ obtained by these two methods are in close agreement. The results from the direct numerical simulation of Eq.~\eqref{eq:RD} with Eq.~\eqref{eq:schnakenberg} are plotted with orange circles and blue crosses for $\epsilon=0.01$ and $\epsilon=0.02$, respectively. 
These results are compared with the right hand side of Eq.~\eqref{eq:phase_response}, which is plotted with the black solid line. The function $\bm{p}(x)$ is set to be $\bm{p}(x)=\transpose{(1,0)}$ in \figref{fig:PRC}(a) and $\bm{p}(x)=\transpose{(\cos \pi x, 0)}$ in \figref{fig:PRC}(b). 
The agreement of the results from the direct numerical simulation and Eq.~\eqref{eq:phase_response} supports the validity of the phase equation \eqref{eq:phase_equation}.
See Appendix~\ref{sec:response_how2measure} for the details of the procedure to determine $\mathrm{PRC}(\phi; \epsilon \bm{p}(x))/\epsilon$ by the direct numerical simulation.

\subsection{Synchronization of a pair of the Schnakenberg systems}
\label{subsec:coupled_systems}
We next investigate the dynamics of a coupled pair of the Schnakenberg systems with delay, which is given as follows:
\begin{align}
    &\pdv{\bm{u}_i(x, t)}{t} 
    =\bm{R}(\bm{u}_i(x,t), \bm{u}_i(x, t-\tau))  \notag\\ 
    &~~~~~~~~~~+ \hat{D} \nabla^2 \bm{u}_i(x, t)+ \epsilon \bm{G}[\bm{u}_j(\cdot,t)](x), \label{eq:coupled_schnakenberg}\\
    &x \in [0,1], ~~~~~i, j=1,2, \notag
\end{align}
where $\bm{R}$ and $\hat{D}$ are given by Eq.~\eqref{eq:schnakenberg_reaction} and Eq.~\eqref{eq:Diffusion_matrix}, respectively, $\epsilon$ is the coupling strength, and $G:C(\Omega, \reals^{N}) \to C(\Omega, \reals^{N})$ describes the coupling between two oscillatory patterns. Simple examples of $\bm{G}$ are local interactions through one component, namely,
\begin{align}
    \bm{G}[\bm{u}(\cdot,t)](x) = 
    \begin{bmatrix}
        u(x, t) \\ 0
    \end{bmatrix}
    \label{eq:u_coupling}
\end{align}
and 
\begin{align}
    \bm{G}[\bm{u}(\cdot,t)](x) = 
    \begin{bmatrix}
        0 \\  v(x, t)
    \end{bmatrix}.
    \label{eq:v_coupling}
\end{align}
Another simple example is
\begin{align}
    \bm{G}[\bm{u}(\cdot,t)](x) = \bm{u}(x, t),
    \label{eq:direct_coupling}
\end{align}
which we refer to as the \emph{direct coupling}.
We may also consider non-local coupling such as
\begin{align}
    \bm{G}[\bm{u}(\cdot, t)](x) = \int_0^{1} \hat{A}(x, x')\bm{u}(x', t) \dd x',
    \label{eq:filtered_coupling}
\end{align}
where $\hat{A}(x, x')$ is a $2 \times 2$ matrix-valued function that determines the influence of the input at $x'$ on the system at $x$.

By approximating $\bm{u}_j(x,t)$ in the coupling term by $\bm{\chi}_{\phi_j(t)}(x)$, where $\phi_j(t)$ is the phase of $j$th system at time $t$, we obtain the coupled phase equations
 \begin{align}
    \hspace{-0.2em}\dv{\phi_i(t)}{t} = \omega + \epsilon \sip{\bm{Q}_{\phi_i(t)}(x)}{\bm{G}[\bm{\chi}_{\phi_j(t)}](x)}. \label{eq:coupled_phase_equation}
 \end{align}

 When $\epsilon$ is sufficiently small, we can perform the averaging approximation~\cite{guckenheimer,Sanders2007, Kuramoto1984} to further simplify Eq.~\eqref{eq:coupled_phase_equation} as follows:
 \begin{align}
    \dv{\phi_i(t)}{t} = \omega + \epsilon \Gamma(\phi_i(t) - \phi_j(t)), \label{eq:averaged_phase_equation}
 \end{align}
where
\begin{align}
    \Gamma(\psi) = \frac{1}{2 \pi}\int_{0}^{2\pi} \sip{\bm{Q}_{\psi + \zeta}(x)}{\bm{G}[\bm{\chi}_{\zeta}](x)} \dd \zeta
\end{align}
is the phase coupling function. 
Then, the dynamics of the phase difference $\psi \coloneqq \phi_1 - \phi_2$ is determined solely by the antisymmetric part of $\Gamma$:
\begin{align}
    \dv{\psi(t)}{t} &=  \epsilon\Ga(\psi(t)), \label{eq:phase_difference_dynamics} \\
    \Ga(\psi) &\coloneqq \Gamma(\psi) - \Gamma(-\psi).
\end{align}
Equation \eqref{eq:phase_difference_dynamics} always has fixed points at $\psi=0$ and $\psi=\pm\pi$, and the in-phase solution $\psi=0$ is stable when $\Ga'(0) < 0$, while the anti-phase solution $\psi=\pm\pi$ is stable when $\Ga'(\pm\pi) < 0$. The stability of these fixed points can also be determined by looking at $\Gamma'(\psi)$ because $\Ga'(0)=2\Gamma'(0)$ and $\Ga'(\pm\pi)=2\Gamma'(\pm\pi)$ hold for any periodic function $\Gamma(\psi)$.

\begin{figure}[]
    \centering
     \includegraphics{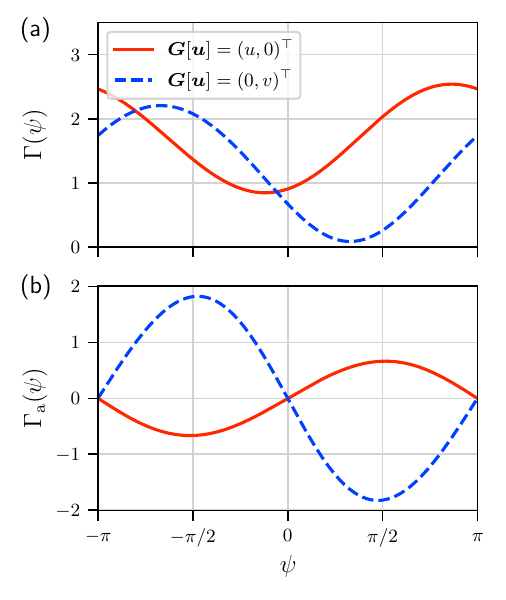}
    \caption{The panels (a) and (b) represent the phase coupling function $\Gamma$ and its  antisymmetric part $\Ga$, respectively.
    In both panels, the orange curves represent the case of coupling via the $u$ component, while the blue dashed curve represents the case of coupling via the $v$ component.
    The former curve indicates anti-phase synchronization, while the latter curve indicates in-phase synchronization.
    }
    \label{fig:Gamma_direct_u_xor_v}
\end{figure}

Figure \ref{fig:Gamma_direct_u_xor_v} shows how the coupling scheme influences the stability of the in-phase and anti-phase solutions. The functions $\Gamma(\psi)$ and $\Ga(\psi)$ are plotted in Figs.~\ref{fig:Gamma_direct_u_xor_v} (a) and (b), respectively. The orange solid and blue dashed curves represent the case of Eq.~\eqref{eq:u_coupling} and the case of Eq.~\eqref{eq:v_coupling}, respectively. The former curve indicates the stable anti-phase solution, while the latter indicates the stable in-phase solution.

\begin{figure}[]
    \centering
    \includegraphics[width=\columnwidth]{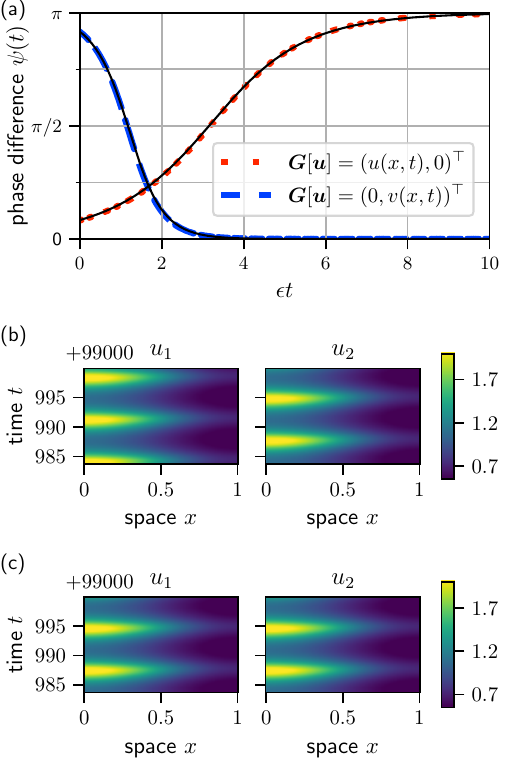}
    \caption{The phase equation can predict the dynamics of the phase difference between the two Schnakenberg systems with delay; coupling via $u$ induces anti-phase synchronization, while coupling via $v$ induces in-phase synchronization. (a) Time series of the phase difference $\psi$. The orange dotted curve corresponds to the coupling via $u$ component, while the blue dashed curve represent the coupling via the $v$ component. (b, c) The spatiotemporal pattern of the $u$ component of $\bm{u}_1$ and $\bm{u}_2$. The coupling scheme is $\bm{G}[\bm{u}(\cdot,t)](x)=\transpose{(u(x,t), 0)}$ in (b) and $\bm{G}[\bm{u}(\cdot,t)](x)=\transpose{(0,v(x,t))}$ in (c). The coupling strength is $\epsilon=0.0001$.
    }
    \label{fig:phase_diff_direct_coupling}
\end{figure}

The above theoretical prediction is confirmed by numerical simulation of  Eq.~\eqref{eq:coupled_schnakenberg} and Eq.~\eqref{eq:phase_difference_dynamics}. 
The red dotted and blue dashed curves in \figref{fig:phase_diff_direct_coupling}(a) represent the time series of $\psi$ obtained by integrating Eq.~\eqref{eq:coupled_schnakenberg} for $u$- and $v$- coupled cases, respectively. 
The curves agree with the predictions from Eq.~\eqref{eq:phase_difference_dynamics}, which are plotted with black solid lines.
As expected, the value of $\psi$ converges to $\pi$ when the systems are coupled via $u$, while it converges to $0$ when the systems are coupled via $v$.
We also show the spatial-temporal patterns of the $u$-components after initial transients for $u$-coupled and $v$-coupled cases in Figs.~\ref{fig:phase_diff_direct_coupling}(b) and (c), respectively.

\section{Optimization for synchronization}
\label{sec:optimal_filter}

Finally, as a demonstration of the utility of the proposed phase reduction method, we optimize the coupling $\bm{G}$ for synchronization by using the method proposed in Kawamura et al.~\cite{Kawamura2017}. Note that, while Eq.~\eqref{eq:coupled_schnakenberg} involves delay, the phase equation \eqref{eq:phase_equation} has exactly the same form as that derived from reaction-diffusion systems without delay. This enables applying theories on synchronization of reaction-diffusion systems to Eq.~\eqref{eq:coupled_schnakenberg}.

\begin{figure}[ht]
    \centering
    \includegraphics[width=\columnwidth]{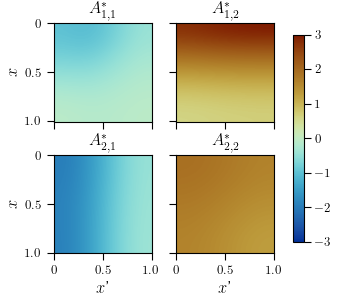}
    \caption{Each component of the optimal filter $A^*(x, x')$, which is given by Eq.~\eqref{eq:optimal_filter}. The normalization constants are $P\simeq 7.54$ and $c(P)\simeq 1.61$.
    }
    \label{fig:optimial_filter}
\end{figure}

Let us consider $\bm{G}[\bm{u}]$ of the form of Eq.~\eqref{eq:filtered_coupling}
with the constraint
\begin{align}
    \hspace{-0.1em}\int_0^{1}\int_0^{1} \sum_{n=1}^2\sum_{m=1}^2 \left|A_{n, m}(x, x')\right|^2\dd x \dd x'= P,    \label{eq:filter_constraint}
\end{align}
where $P> 0$ is a normalization constant and $A_{n,m}(x, x')$ is the $(n,m)$ component of the $2 \times 2$ matrix-valued function $\hat{A}(x, x')$.
Then the linear stability of in-phase synchronization, $\psi=0$, is maximized when
\begin{align}
   A_{n,m}(x, x')&= 
   \frac{c(P)}{2\pi} \int_{0}^{2\pi} Q_n(x,\zeta) U_m(x', \zeta) \dd \zeta 
    \notag \\
    &\eqqcolon A_{n,m}^*(x, x'),
    \label{eq:optimal_filter}
\end{align}
where $c(P)$ is a normalization constant to satisfy Eq.~\eqref{eq:filter_constraint}, $Q_n(x, \zeta)$ is the $n$th component of the phase sensitivity function $\bm{Q}_\zeta(x)$, which is defined by Eq.~\eqref{eq:Q_phi}, and $U_m(x', \zeta)$ is the $m$th component of $\bm{U}_{\zeta}(x)$, which is defined by Eq.~\eqref{eq:U_phi}~\cite{Kawamura2017}. Figure \ref{fig:optimial_filter} shows each component of the optimal filter $\hat{A}^*$. We observe that $A_{1,2}^*(x, x')$ for $x \simeq 0$ and $A_{2,1}^*(x, x')$ for $x'\simeq 0$ have relatively large absolute values. This observation implies the importance of the $u$ component at around $x=0$ for maximizing the stability.

\begin{figure}[ht]
    \centering 
    \includegraphics[width=\columnwidth]{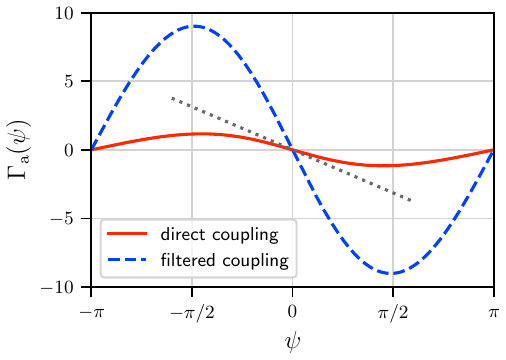}
    \caption{ The antisymmetric part of the phase coupling function. The case of direct coupling is plotted with the orange solid curve, while the case of coupling via optimal filter is plotted with the blue dashed curve. The gray dotted line is a visual guide with slope $-2$.
    Note that $\Ga'(0) \simeq -1.35$ is larger than $-2$.
    The normalization constant $P$ is set as $P\simeq 7.54$.
        }
    \label{fig:coupling_function_filtered}
\end{figure}

The blue dashed line in \figref{fig:coupling_function_filtered} represents $\Ga(\psi)$ for the case of coupling with the optimal filter. Its slope at $\psi=0$ is negative, implying that in-phase synchronization is stable. 
To illustrate the efficiency of the optimal filter, we compare it with $\Ga(\psi)$ in the direct coupling case. 
For a fair comparison, the constant $P$ is chosen so that $\frac{1}{2 \pi}\int_{0}^{2\pi} \norm{\bm{G}\left( \bm{\chi_\phi}(x) \right)}_2^2 \dd \phi$ is the same as the filtered coupling, where $\norm{\cdots}_2$ represents the $L_2$ norm of a given column vector. The slope of $\Ga(\psi)$ at $\psi = 0$ in the filtered coupling case is steeper when compared to the direct coupling case, implying that the local stability of in-phase synchronization is increased by using the optimal filter.

The gray dotted line in \figref{fig:coupling_function_filtered} is a visual guide with the slope of $-2$. It is known that the direct coupling yields $\Ga'(0)=2\Gamma'(0)=-2$ in the absence of delay~\cite{Kawamura2017}. 
This formula holds because $-\Gamma'(0)$ reduces to the inner product of the phase sensitivity function and the zeroth Floquet mode, which is $1$ owing to the normalization condition of the phase sensitivity function.
Similarly, in the presence of delay, $-\Gamma'(0)$  for the direct coupling case reduces to the inner product $\sip{\bm{Q}_{\phi(t)}(\bm{r})}{\bm{U}(\bm{r}, t)/\omega}$. 
This quantity, however, is not necessarily $1$ because the normalization condition \eqref{eq:normalization} involves the bilinear form $\blf{\cdots}{\cdots}$, not the inner product $\sip{\cdots}{\cdots}$. 
A comparison of Eq.~\eqref{eq:blf}, which defines the bilinear form $\blf{\cdots}{\cdots}$, with Eq.~\eqref{eq:sip}, which defines the inner product $\sip{\cdots}{\cdots}$, shows that they differ in whether the integral with respect to $\sigma$ from $-\tau$ to $0$ is included.
Thus, $\Ga'(0)=-2$ for the direct coupling case is no longer guaranteed when the evolution of the system depends on the past.
Indeed, the orange solid curve in \figref{fig:coupling_function_filtered}, representing $\Ga(\psi)$ for the direct coupling case, shows $\Ga'(0)\simeq-1.35 > -2$.

\begin{figure}[ht]
    \centering
    \includegraphics{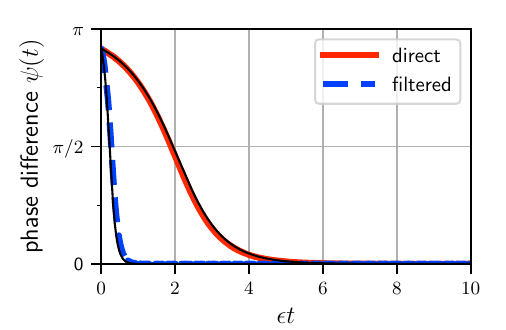}
    \caption{Comparison of the time series of the phase difference for direct and filtered coupling. The orange solid curve is for the case of direct coupling, while the blue dashed curve is for the case of coupling via optimal filter.
    The thin black curves represent the prediction from the phase equation \eqref{eq:averaged_phase_equation}. The coupling strength is $\epsilon=0.0001$.
    }
    \label{fig:phase_diff_optimized}
\end{figure}

The optimal filter not only increases the local stability of in-phase synchronization but also shortens the time required for the oscillatory patterns with relatively large initial phase difference to synchronize. 
The blue dashed curve in \figref{fig:phase_diff_optimized} shows the time series of the phase difference of the system \eqref{eq:coupled_schnakenberg} with optimal filtered coupling. 
The phase difference decays from the initial value $11 \pi/12$ to $0$ faster than the case of direct coupling, plotted by red solid curve. 
Thus, optimization of the interaction based on the phase reduction theory stabilizes and accelerates in-phase synchronization of the coupled Schnakenberg systems with delay.

\section{Concluding remarks}
\label{sec:concluding_remarks}
In this study, we developed a phase reduction method tailored to reaction-diffusion systems with a discrete delay. Adopting the adjoint-equation based approach, which has been known to be effective for infinite-dimensional systems, our formulation extends the phase reduction method to incorporate both spatial degrees of freedom and delay. The resulting phase equation was derived using an appropriate bilinear form.
The proposed method was numerically validated by using the delayed Schnakenberg system in one-spatial dimension as an example.
Specifically, we confirmed that the phase response curves obtained from the phase equation agreed well with those obtained from the direct numerical simulation of the original system.
We also showed that, when such a system was interacting with another oscillatory pattern, the modulation of the rhythms of the two systems was well captured by the corresponding phase equations.

We obtained the phase sensitivity function as a periodic solution of the adjoint equation. Alternatively, the phase sensitivity function may be defined as the functional derivative of a phase functional, as was done for reaction-diffusion systems without delay~\cite{Nakao2014}. 
While proving the equivalence of the two phase sensitivities obtained through these two approaches is beyond the scope of this study, such a proof will lead to a more solid understanding of the modulation of rhythms of the oscillatory patterns. 
In Appendix \ref{sec:equivalence}, we discussed the relation between the phase sensitivity function in the present study and the functional derivative of the asymptotic phase, assuming that these two functions were well defined.

We obtained the adjoint equation analytically and discretized it to solve numerically. Such an approach is referred to as the \emph{discretization of the adjoint} (DA)~\cite{Luchini2014}. The DA approach is distinguished from the alternative approach known as the \emph{adjoint of discretization} (AD), which involves first carrying out the discretization of a given system and then obtaining the adjoint equation of the resultant discrete system~\cite{Luchini2014}. 
While the AD approach is also applicable to delayed PDEs~\cite{Skene2022}, the DA approach often helps elucidating mathematical properties of the adjoint equation, such as symmetries and conditions to be satisfied~\cite{Kawamura2013, Kawamura2014,Kawamura2018,Kawamura2019,Kawamura2022}. 
In the case of the delayed reaction-diffusion equation \eqref{eq:RD} with the boundary condition \eqref{eq:zero_neumann}, we obtained the boundary condition \eqref{eq:zero_neumann_adjoint} of the adjoint equation~\eqref{eq:adjoint}.

In Appendix~\ref{sec:adjoint}, we claimed that the periodic solution $\bm{Q}^{(t)}(\sigma)(\bm{r})$ extracts only the phase mode because Eq.~\eqref{eq:orthogonal} holds for any nonzero Floquet mode $\bm{v}_{\lambda}^{(t)}(\bm{\sigma})(\bm{r})$. However, the completeness of the set of Floquet solutions is not guaranteed in delayed systems~\cite{Hale1993}.
Nonetheless, our numerical simulations in Sec.~\ref{sec:numerical_validation} indicate that the present theory is valid, at least for certain delayed reaction-diffusion systems.
While the present theory is expected to be applicable to a range of systems, the precise condition for its validity should be clarified by establishing a Floquet theory for delayed reaction-diffusion equations.

There are several promising directions for extending the present work.
First, delay may appear in the diffusion term in some systems~\cite{Hale1994, Wu1996}, and extending the theory to deal with such systems will widen the range of its application.
Second, the phase reduction method may be extended to partial differential equations involving distributed delays by appropriately modifying the bilinear form~\eqref{eq:blf}.
Finally, developing a theory for delay partial differential equations coupled with constraints will facilitate the investigation on synchronization in climate systems~\cite{Cessi2000}.

\begin{acknowledgments}
    This study was supported by JSPS KAKENHI Grant No. JP24K23907 to A. O. and JSPS KAKENHI Grant Numbers JP24K06910 and JP25K01160 to Y. K.
    Some of the numerical simulations were conducted using the Earth Simulator at JAMSTEC.
\end{acknowledgments}

\appendix
\section{Derivation of the adjoint equation \eqref{eq:adjoint}}
\label{sec:adjoint}
In this appendix, we obtain the adjoint system \eqref{eq:adjoint} with the boundary condition \eqref{eq:zero_neumann_adjoint} from Eq.~\eqref{eq:constant_blf}.

From the definition \eqref{eq:blf}, 
\begin{align}
    & \dv{}{t}\blf{\bm{q}^{(t)}(\sigma)(\bm{r})}{\bm{v}^{(t)}(\sigma)(\bm{r});t} \notag\\
    & = A(t) + B(t), \label{eq:A+B}
\end{align}
where
\begin{widetext}
    \begin{align}
        &A(t)= \dv{}{t}
            \sip{\bm{q}(\bm{r},t)}{\bm{v}(\bm{r},t)}=\sip{\pdv{\bm{q}(\bm{r},t)}{t}}{\bm{v}(\bm{r},t)} +\sip{\bm{q}(\bm{r},t)}{\pdv{\bm{v}(\bm{r},t)}{t}}, \label{eq:A(t)}\\
        &B(t)= \dv{}{t}\int_{-\tau}^{0} 
                \sip{\bm{q}(\bm{r}, t+\tau+\sigma)}
                    {\hat{R}_2(\bm{r}, t+\tau+\sigma)\bm{v}(\bm{r}, t+\sigma)}
            \dd \sigma.
    \end{align}
\end{widetext}
Note that, for any matrix-valued function $\hat{M}(\bm{r}, t)$,
\begin{align}
    \sip{\bm{q}(\bm{r}, t)}{\hat{M}(\bm{r},t)\bm{v}(\bm{r}, t)} 
    &=
    \int_{\Omega} \bm{q}(\bm{r}, t)\hat{M}(\bm{r},t)\bm{v}(\bm{r}, t) \dd \bm{r} \notag \\
    &=\sip{\bm{q}(\bm{r}, t)\hat{M}(\bm{r},t)}{\bm{v}(\bm{r}, t)}.
    \label{eq:adjoint_matrix}
\end{align}
From Eqs.~\eqref{eq:A(t)}, \eqref{eq:linearized}, \eqref{eq:define_L}, and \eqref{eq:adjoint_matrix}, we obtain
\begin{align}
    A(t)
    &=\sip{\pdv{\bm{q}(\bm{r},t)}{t}}{\bm{v}(\bm{r},t)}
    +\sip{\bm{q}(\bm{r}, t)\hat{R}_1(\bm{r},t)}{\bm{v}(\bm{r}, t)} \notag\\
    &~~~~+\sip{\bm{q}(\bm{r}, t)\hat{R}_2(\bm{r},t)}{\bm{v}(\bm{r}, t-\tau)} \notag \\
    &~~~~+\sip{\bm{q}(\bm{r},t)\hat{D}}{\nabla^2\bm{v}(\bm{r}, t)}.
    \label{eq:A_flipped}
\end{align}

To further proceed, we use Green's second identity to obtain
\begin{align}
    \sip{\bm{q}(\bm{r},t)\hat{D}}{\nabla^2\bm{v}(\bm{r}, t)}
    = \sip{\nabla^2\bm{q}(\bm{r},t)\hat{D}}{\bm{v}(\bm{r}, t)} + S(t),
        \label{eq:adjoint_nabla2}
\end{align}
where 
\begin{align}
    S(t) = \sum_{n=1}^{N} 
        \int_{\partial \Omega} 
            \mathbf{n}\cdot
            &\left[ 
                \tilde{q}_n(\bm{r},t) \nabla v_n(\bm{r},t) \right.\notag \\ 
                &\left.
                ~~-v_n(\bm{r},t) \nabla \tilde{q}_n(\bm{r},t)
            \right] \dd \partial \Omega
\end{align}
and $\tilde{q}_n(\bm{r},t)$ is the $n$th component of the vector $\bm{q}(\bm{r},t)\hat{D}$.
From the condition \eqref{eq:zero_neumann_linarized}, $S(t)$ simplifies to
\begin{align}
    S(t) = -\sum_{n=1}^{N} 
    \int_{\partial \Omega} 
    \mathbf{n}\cdot
    \left[
    v_n(\bm{r},t) \nabla \tilde{q}_n(\bm{r},t)\right]
    \dd \partial \Omega.
    \label{eq:S(t)}
\end{align}

The second term $B(t)$ of Eq.~\eqref{eq:A+B} can be written as follows:
\begin{widetext}
\begin{align}
    B(t)
    &= \dv{}{t}
    \int_{-\tau}^{0} \sip{\bm{q}(\bm{r}, t+\tau+\sigma)\hat{R}_2(\bm{r}, t+\tau+\sigma)}
            {\bm{v}(\bm{r}, t+\sigma)}
    \dd \sigma \notag \\
    &=\int_{-\tau}^{0} 
    \dv{}{t}
        \sip{\bm{q}(\bm{r}, t+\tau+\sigma)\hat{R}_2(\bm{r}, t+\tau+\sigma)}
            {\bm{v}(\bm{r}, t+\sigma)}
    \dd \sigma \notag \\
    &=\int_{-\tau}^{0} 
    \dv{}{\sigma}
        \sip{\bm{q}(\bm{r}, t+\tau+\sigma)\hat{R}_2(\bm{r}, t+\tau+\sigma)}
            {\bm{v}(\bm{r}, t+\sigma)}
    \dd \sigma \notag \\
    &= \sip{\bm{q}(\bm{r}, t+\tau)\hat{R}_2(\bm{r}, t+\tau)}
            {\bm{v}(\bm{r}, t)}-
        \sip{\bm{q}(\bm{r}, t)\hat{R}_2(\bm{r}, t)}
            {\bm{v}(\bm{r}, t- \tau)}.
    \label{eq:B_integrated}
\end{align}
\end{widetext}

Inserting \eqref{eq:adjoint_nabla2} into \eqref{eq:A_flipped} and then the resultant equation together with \eqref{eq:B_integrated} into Eq.~\eqref{eq:A+B}, we have
\begin{align}
    &\dv{}{t}\blf{\bm{q}^{(t)}(\sigma)(\bm{r})}{\bm{v}^{(t)}(\sigma)(\bm{r});t} \notag\\
    &=\sip{\pdv{\bm{q}(\bm{r},t)}{t}}{\bm{v}(\bm{r},t)} \notag \\
    &~~~+\sip{\bm{q}(\bm{r}, t)\hat{R}_1(\bm{r},t)}{\bm{v}(\bm{r}, t)} \notag\\
    &~~~+\sip{\bm{q}(\bm{r}, t+\tau)\hat{R}_2(\bm{r},t+\tau)}{\bm{v}(\bm{r}, t)} \notag \\
    &~~~+\sip{\nabla^2\bm{q}(\bm{r},t)\hat{D}}{\bm{v}(\bm{r}, t)} + S(t).
\end{align}
The condition \eqref{eq:constant_blf} is satisfied for any $\bm{v}(\bm{r}, t)$ when Eq.~\eqref{eq:adjoint} holds and $S(t) = 0$.
From Eq.~\eqref{eq:S(t)} it follows that $\bm{q}(\bm{r}, t)$ must satisfy the zero Neumann boundary condition \eqref{eq:zero_neumann_adjoint} for $S(t)$ to vanish for any $\bm{v}(\bm{r},t)$.
Thus, we obtain Eq.~\eqref{eq:adjoint} with the boundary condition \eqref{eq:zero_neumann_adjoint}.

\section{Derivation of Equation \eqref{eq:orthogonal}}
\label{sec:zero_eigenfunction}
Here we derive Eq.~\eqref{eq:orthogonal}. The idea is the same as Ref.~\cite{Novicenko2012}.
The condition \eqref{eq:constant_blf} implies
\begin{align}
    &\blf{\bm{Q}^{(t+T)}(\sigma)(\bm{r})}{\bm{v}^{(t+T)}_\lambda(\sigma)(\bm{r}); t+T} \notag\\
    &=\blf{\bm{Q}^{(t)}(\sigma)(\bm{r})}{\bm{v}^{(t)}_\lambda(\sigma)(\bm{r}); t}
.\label{eq:Q_v_period}
\end{align}
Because $\bm{Q}^{(t)}(\sigma)(\bm{r})$ and $\hat{R}_2(\bm{r},t)$ are $T$-periodic with respect to $t$ and $\bm{v}_\lambda^{(t)}$ is assumed to have the form of Eq.~\eqref{eq:v_lambda}, we have
\begin{align}
    &\blf{\bm{Q}^{(t+T)}(\sigma)(\bm{r})}{\bm{v}^{(t+T)}_\lambda(\sigma)(\bm{r}); t+T} \notag\\
    &=\blf{\bm{Q}^{(t)}(\sigma)(\bm{r})}{e^{\lambda (t+T)} \bm{p}_{\lambda}(\bm{r}, t);t} \notag \\
    &=e^{\lambda T}\blf{\bm{Q}^{(t)}(\sigma)(\bm{r})}{\bm{v}^{(t)}_\lambda(\sigma)(\bm{r}); t}. \label{eq:Q_v_period_2}
\end{align}
Comparing Eq.~\eqref{eq:Q_v_period} and \eqref{eq:Q_v_period_2}, we obtain Eq.~\eqref{eq:orthogonal} for $\lambda \neq 0$.

\section{Numerical procedure to obtain phase sensitivity function}
\label{sec:response_how2measure}
To integrate Eq.~\eqref{eq:RD} with Eq.~\eqref{eq:schnakenberg}, we discretized the space and time as $x_l = l \Delta x~(l=0,1,\dots,L)$ and $t_k = k \Delta t$ $(k=-K,-K+1,\dots)$, where $\Delta x \coloneqq 1/L$ and $\Delta t \coloneqq \tau/K$ are the grid size and time step, respectively, and we set $L=128$ and $K=400$. 
Following Lee~\emph{et al}.~\cite{SeirinLee2010}, we adopted the fully implicit treatment of the diffusion term and explicit treatment of the reaction term.
Specifically, the discretized dynamics is given as 
\begin{align}
     &\frac{\bm{u}(x_l,t_{k+1}) - \bm{u}(x_l,t_{k})}{\Delta t}  \notag \\
     &=  \bm{R}(\bm{u}(x_l,t_k), \bm{u}(x_l, t_{k-K}))  
     + \hat{D}\frac{1}{\Delta x^2}\left[\bm{u}(x_{l+1},t_{k+1}) \right.\notag \\
     &~~~~~~~~~ \left.  + \bm{u}(x_{l-1}, t_{k+1}) - 2 \bm{u}(x_{l}, t_{k+1})\right]
     +\epsilon\bm{p}(x_l, t_k),\label{eq:discretized}\\
    &l=1,\dots,L-1, \notag\\
    &k=0, 1, \dots, \notag
\end{align}
For $l=0$ and $l=L$, the dynamics is the same as Eq.~\eqref{eq:discretized} 
except that the diffusion term is replaced with
\begin{align}
     \hat{D}\frac{1}{\Delta x^2}\left[ 2 \bm{u}(x_1,t_{k+1}) - 2 \bm{u}(x_0,t_{k+1})\right]
     \label{eq:discretized_left}
\end{align}
and
\begin{align}
    \hat{D}\frac{1}{\Delta x^2}\left[ 
        2\bm{u}(x_{L-1},t_{k+1}) - 2 \bm{u}(x_{L}, t_{k+1})
     \right],
     \label{eq:discretized_right}
\end{align}
respectively, to reflect the boundary condition \eqref{eq:zero_neumann};
See Chap.~5 of Linge~\emph{et al.}~\cite{Linge2016} for a discussion on the implementation of the boundary condition.
The initial condition was given by
\begin{align}
    \bm{u}(x_l,t_{k}) = 
    \begin{bmatrix}
        1 - 0.1 \cos (l \pi/L) \\
        1 - 0.1 \cos (l \pi/L)
    \end{bmatrix},\\
    k=-K, -K+1,\dots,0, \notag\\
    l = 0,1,\dots, L. \notag
\end{align}
We obtained  $\left\{ \bm{u}(x_l, t_{k+1}) \right\}_{l=0}^{L}$ from $\left\{ \bm{u}(x_l, t_{k}) \right\}_{l=0}^{L}$ by solving Eqs. \eqref{eq:discretized}, \eqref{eq:discretized_left}, and \eqref{eq:discretized_right} simultaneously.

The period of the oscillatory solution was calculated from the time series of the $u$ component at $x=0$. 
Specifically, a cycle was detected when the value of $\bm{u}(0,t)$ crossed $1.5$. 
This choice of the Poincare section is motivated by Arai~\emph{et al.}~\cite{Arai2025}, 
which proposed using measurements from the region where the amplitude of the phase sensitivity function is maximized.
We evaluated the time $t$ at which $\bm{u}(0,t) = 1.5$ by linearly interpolating the simulation data, and the system was considered to have converged to a limit cycle when the difference between the periods of two succeeding oscillations became smaller than $10^{-6}$.

To measure the phase response curve, $\mathrm{PRC}(\phi; \epsilon \bm{p}(x))$, we prepared a system that converged to the limit cycle and applied the perturbation $\epsilon \bm{p}(x)$ when the phase is $\phi$.
We then waited until the system converged to the limit-cycle again, using the same convergence criteria described above.
Finally, $\mathrm{PRC}(\phi; \epsilon \bm{p}(x))$ was obtained by subtracting the phase of the unperturbed system from that of the perturbed system.

\section{The relation to the asymptotic phase}
\label{sec:equivalence}
In Subsec.~\ref{subsec:adjoint}, we obtained the phase equation \eqref{eq:phase_equation} by 
quantifying the response of the phase to perturbation by using the bilinear form 
\eqref{eq:blf}. 
An alternative approach for obtaining the phase equation may be to 
define the phase sensitivity to be the functional derivative of the asymptotic phase evaluated on the limit cycle, as was done for systems without delay~\cite{Nakao2014}.
Below, we discuss the relation between these two approaches in the case of Eq.~\eqref{eq:RD}.

\subsection{A recapitulation of the asymptotic phase functional for reaction-diffusion systems without delay}
\label{sec:nakao_et_al_2014}
For reaction-diffusion systems without delay, Nakao~\emph{et al.}~\cite{Nakao2014} defined the phase functional $\Theta\left[ \bm{X}(\bm{r}) \right]$ of a spatial pattern $\bm{X}(\bm{r})$ as follows. Given a limit cycle solution $\bm{X}_0(\bm{r}; \theta)$ of the unperturbed system parameterized by the phase variable $\theta$ that evolves with frequency $\omega$ as $\theta(t) = \omega t \mod 2 \pi$, the authors assigned the phase $\theta(t)$ to the set of states of the unperturbed system, $\{\bm{X}(\bm{r}, t)\}$, that satisfies
\begin{align}
    \lim_{t\to+\infty} \norm{\bm{X}(\bm{r},t) - \bm{X_0}\bm{(}\bm{r}; \theta(t)\bm{)}} = 0,
\end{align}
where $\norm{\cdots}$ represents the $L^2$ norm of a spatial pattern.
Then the functional $\Theta\left[ \bm{X}(\bm{r}) \right]$ was defined to satisfy
\begin{align}
    \Theta \left[ \bm{X}(\bm{r},t) \right]  = \theta(t).
\end{align}

Furthermore, the phase sensitivity at the phase $\theta$, denoted by $\bm{Q}(\bm{r}; \theta)$, was defined to be the functional derivative of $\Theta(\bm{X})$ at the point on the limit cycle with phase $\theta$~\footnote{While in Ref.~\cite{Nakao2014}, the authors referred to $\frac{\delta \Theta\left[  \bm{X}(\bm{r})\right]}{\delta \bm{X}(\bm{r})}$ as the \emph{functional gradient} instead of \emph{functional derivative}, we adopt the latter in this article.}, i.e., 
\begin{align}
    \bm{Q}(\bm{r}; \theta) = 
    \left.\frac{\delta \Theta\left[  \bm{X}(\bm{r})\right]}{\delta \bm{X}(\bm{r})}\right|_{\bm{X}(\bm{r}) = \bm{X}_0(\bm{r}; \theta)}.
    \label{eq:Q_derivative_RD}
\end{align}
It was shown that $\bm{Q}(\bm{r}; \theta)$ is equivalent to the zeroth Floquet mode of an adjoint equation of the linearization around the limit-cycle solution.
See Appendix B of Ref.~\cite{Nakao2014} for more details.

\subsection{The asymptotic phase functional for reaction-diffusion systems with delay}
As an analogy, we may also assign the asymptotic phase $\phi$ to the elements of $\mathcal{C}$ and introduce the phase functional $\Phi[\bm{u}(\sigma)(\bm{r})]:\mathcal{C} \to \reals$ such that $\Phi\bm{(}\bm{u}^{(t)}(\sigma)(\bm{r})\bm{)} = \phi(t)$, where $\phi(t)$ evolves as $\phi = \omega t$.

Now we discuss the relation between $\bm{Q}^{(t)}(\sigma)(\bm{r})$, defined by Eq.~\eqref{eq:qt_history}, and the functional derivative of $\Phi[\bm{u}(\sigma)(\bm{r})]$.
Suppose the variation of the functional $\Phi[\bm{u}(\sigma)(\bm{r})]$ is well-defined on the limit cycle. 
More specifically, assume 
\begin{align}
    &\Phi[\bm{\chi}^{(t)}(\sigma)(\bm{r}) + \bm{v}(\sigma)(\bm{r})] \notag \\
    &= \phi 
    + 
    \int_{-\tau}^{0} \int_\Omega
        \left.\frac{\delta \Phi[ \bm{u} ]}{\delta \bm{u}}\right|_{\bm{u}= \bm{\chi}^{(t)}(\sigma)(\bm{r})}\bm{v}(\sigma)(\bm{r})  \dd \bm{r} \dd \sigma  \notag \\
    &~~~~~~+ o(\norm{\bm{v}(\sigma)(\bm{r})}_\mathrm{sup}),
    \label{eq:phi_variation}
\end{align}
where $\norm{\bm{v}(\sigma)(\bm{r})}_\mathrm{sup}$ is the supremum norm of $\bm{v}(\sigma)(\bm{r})$, i.e., $\norm{\bm{v}(\sigma)(\bm{r})}_\mathrm{sup}\coloneqq \sup_{\sigma \in [-\tau, 0]}\norm{\bm{v}(\sigma)(\bm{r})}$.
As in  Sec.~\ref{sec:phase_reduction_theory}, $\bm{\chi}^{(t)}(\sigma)(\bm{r}) \coloneqq \bm{\chi}(\bm{r}, t+\sigma)$ is the limit cycle solution of the unperturbed system.
Also, let $\bm{u}(\bm{r}, t)$ be a solution of Eq.~\eqref{eq:RD} for $\epsilon=0$ 
and define $\bm{u}^{(t)}(\sigma)(\bm{r}) \coloneqq \bm{u}(\bm{r}, t+\sigma)$.
Since the difference between the asymptotic phases of $\bm{u}^{(t)}(\sigma)(\bm{r})$ and $\bm{\chi}^{(t)}(\sigma)(\bm{r})$ is constant with respect to time in the absence of perturbation, we have
\begin{align}
    \dv{}{t}\left[\Phi[\bm{u}^{(t)}(\sigma)(\bm{r})] - \Phi[\bm{\chi}^{(t)}(\sigma)(\bm{r})] \right] = 0.
    \label{eq:phase_diff_constant}
\end{align}

Note that Eq.~\eqref{eq:phi_variation} yields
\begin{align}
    &\Phi[\bm{u}^{(t)}(\sigma)(\bm{r})] - \Phi[\bm{\chi}^{(t)}(\sigma)(\bm{r})]\notag \\
    &\hspace{-1em}= \int_{-\tau}^{0} \int_\Omega
    \left.\frac{\delta \Phi[ \bm{u}]}{\delta \bm{u}}\right|_{\bm{u}=\bm{\chi}^{(t)}(\sigma)(\bm{r})}\bm{v}^{(t)}(\sigma)(\bm{r})
    \dd \sigma \dd \bm{r},
        \label{eq:phase_diff_as_variation}
\end{align}
to the lowest order of $\norm{\bm{v}^{(t)}(\sigma)(\bm{r})}_\mathrm{sup}$, where $\bm{v}^{(t)}(\sigma)(\bm{r}) \coloneqq \bm{v}(\bm{r}, t+\sigma)$ and $ \bm{v}(\bm{r}, t)= \bm{u}\bm({r}, t) - \bm{\chi}(\bm{r}, t)$. 
Equations \eqref{eq:phase_diff_constant} and \eqref{eq:phase_diff_as_variation} imply
\begin{align}
    \dv{}{t}
    \left[
        \int_{-\tau}^{0} \int_\Omega
    \left.\frac{\delta \Phi[ \bm{u}]}{\delta \bm{u}}\right|_{\bm{u}= \bm{\chi}^{(t)}(\sigma)(\bm{r})}\bm{v}^{(t)}(\sigma)(\bm{r})
    \dd \sigma \dd \bm{r}
        \right] = 0.
        \label{eq:constant_variation}
\end{align}

Provided that $\bm{v}(\bm{r}, t)$ obeys the linearized equation \eqref{eq:linearized}, 
Eq.~\eqref{eq:constant_variation} is satisfied when 
\begin{align}
     &\left.
        \frac{\delta \Phi[ \bm{u}]}
                {\delta \bm{u}}
      \right|_{
       \bm{u}= \bm{\chi}^{(t)}(\sigma)(\bm{r})} 
       \notag\\
     &=\bm{Q}^{(t)}(\sigma)(\bm{r})\{\delta(\sigma)+\hat{\hat{R}}_2(\bm{r},t + \tau + \sigma)\},
     \label{eq:variation_q}
\end{align}
where we defined the delta function such that 
\begin{align}
    \int_{-\tau}^{0} f(\sigma)\delta(\sigma) \dd \sigma = f(0).
\end{align}
This can be shown by inserting Eq.~\eqref{eq:variation_q} into the left hand side of Eq.~\eqref{eq:constant_variation} and noting that, from Eq.~\eqref{eq:constant_blf}, 
\begin{align}
    \dv{}{t}\blf{\bm{Q}^{(t)}(\sigma)(\bm{r})}{\bm{v}^{(t)}(\sigma)(\bm{r})}.
\end{align}
Hence, we conjecture that the functional derivative of $\Phi$ and $\bm{Q}^{(t)}(\sigma)(\bm{r})$ are connected through Eq.~\eqref{eq:variation_q}. 
Equation \eqref{eq:variation_q} implies that the functional derivative of the phase functional is not equivalent to the phase sensitivity function, in contrast to the case without delay, c.f., Eq.~\eqref{eq:Q_derivative_RD}.

\subsection{A remark on the phase response curve obtained in Sec.~\ref{sec:numerical_validation}}
Finally, we remark that the phase response curve, which we obtained numerically in Sec.~\ref{sec:numerical_validation}, provides a link between the asymptotic phase and our phase sensitivity function $\bm{Q}_\phi(x)$.

Let $\bm{\chi}^{(t)}(\sigma)(x)$ be the limit-cycle solution of the unperturbed system and $\bm{u}^{(t)}(\sigma)(x)$ be a solution of the perturbed system.
By definition, the phase response curve $\mathrm{PRC}(\phi(t_0); \epsilon \bm{p}(x))$ is calculated as
\begin{align}
       \mathrm{PRC}(\phi(t_0); \epsilon \bm{p}(x)) \notag \\
       = \Phi\left[\bm{u}^{(t)}(\sigma)(x)\right] - \Phi\left[\bm{\chi}^{(t)}(\sigma)(x)\right] \label{eq:prc_u}
\end{align}
for any $t \geq t_0 + \Delta t$, where $t_0$ is the time at which the perturbation $\epsilon \bm{p}(x)/\Delta t$ is applied and $\Delta t$ is the duration of the perturbation, c.f. Sec.~\ref{sec:numerical_validation}.

When the perturbation strength $\epsilon$ and length of the perturbation $\Delta t$ are sufficiently small, we may approximate
\begin{align}
    {\bm{u}}^{(t)}(\sigma)(x) \simeq \bm{\chi}^{(t)}(\sigma)(x) + \epsilon \bm{p}(\sigma)(x) \label{eq:utilde_approx}
\end{align}
for $t = t_0 + \Delta t$, where 
\begin{align}
    \bm{p}(\sigma)(x) \coloneqq 
    \begin{cases}
        \bm{p}(x) & (\sigma = 0) \\
        0 & (-\tau \leq \sigma < 0)
    \end{cases}.
\end{align}

Inserting Eq.~\eqref{eq:utilde_approx} into Eq.~\eqref{eq:prc_u}, we have
\begin{align}
    &\mathrm{PRC}\bm{(}\phi; \epsilon \bm{p}(x)\bm{)} \notag \\
    &\simeq \Phi\bm{(}
        \bm{\chi}_{\phi}(\sigma)(x) + \epsilon \bm{p}(\sigma)(x)
        \bm{)} - \Phi\bm{(}\bm{\chi}_{\phi}(\sigma)(x)\bm{)},
    \label{eq:prc_phi_approx_chi}
\end{align}
where we assumed that the domain of $\Phi$ can be extended to allow discontinuity at $\sigma=0$.

Comparing Eq.~\eqref{eq:prc_phi_approx_chi} with Eq.~\eqref{eq:phase_response}, we obtain a relation between the asymptotic phase functional $\Phi$ and the phase sensitivity function $\bm{Q}_\phi(x)$ as
\begin{align}
    &\Phi\bm{(}
        \bm{\chi}_{\phi}(\sigma)(x) + \epsilon \bm{p}(\sigma)(x)
        \bm{)}
    - \Phi\bm{(}\bm{\chi}_{\phi}(\sigma)(x)\bm{)} \notag \\
    &\simeq \sip{\bm{Q}_{\phi}(x)}{\epsilon\bm{p}(x)} \notag \\
    & = \blf{\bm{Q}^{(\phi/\omega)}(\sigma)(x)}{\epsilon\bm{p}(\sigma)(x); t},
    \label{eq:relation_PRC_Q}
\end{align}
which is consistent with the conjecture \eqref{eq:variation_q}.

\bibliography{citations}

\end{document}